\documentclass[%
twocolumn, superscriptaddress,
showpacs,preprintnumbers,
nofootinbib,
 amsmath,amssymb,
 aps,
pra,
]{revtex4}

\usepackage{graphicx}
\usepackage{amsfonts}
\usepackage{dcolumn}
\usepackage{bm}
\usepackage{hyperref}

\newcommand{\beq}{\begin{equation}}
\newcommand{\eeq}{\end{equation}}
\newcommand{\beqa}{\begin{eqnarray}}
\newcommand{\eeqa}{\end{eqnarray}}
\newcommand{\ra}{\rangle}

\makeatletter
\newcommand{\doublewidetilde}[1]{{%
  \mathpalette\double@widetilde{#1}%
}}
\newcommand{\double@widetilde}[2]{%
  \sbox\z@{$\m@th#1\widetilde{#2}$}%
  \ht\z@=.9\ht\z@
  \widetilde{\box\z@}%
}
\makeatother

\usepackage{calrsfs}
\DeclareMathAlphabet{\pazocal}{OMS}{zplm}{m}{n}

\begin{document}

\title{Fast and robust particle shuttling for quantum science and technology}

%
\author{L. Qi}
\affiliation{Department of Physics, Massachusetts Institute of Technology, Cambridge, Massachusetts 02139, USA}
\author{J. Chiaverini}
\affiliation{Lincoln Laboratory, Massachusetts Institute of Technology, Lexington, Massachusetts 02421, USA}
\author{H. Espin\'os}
\affiliation{Department of Physical Chemistry, University of the Basque Country UPV/EHU,\\ Apdo 644, 48080 Bilbao, Spain}
\author{M. Palmero}
\affiliation{Department of Applied Physics, University of the Basque Country (UPV/EHU), 48013 Bilbao, Spain}
\author{J. G. Muga}
\affiliation{Department of Physical Chemistry, University of the Basque Country UPV/EHU,\\ Apdo 644, 48080 Bilbao, Spain}
%




\begin{abstract}
  We review  methods to shuttle  quantum particles  fast and robustly.
  Ideal robustness amounts to the invariance of the desired transport results with respect to deviations,
  noisy or otherwise,  from the nominal driving protocol for the control parameters; this can include environmental perturbations.  ``Fast'' is defined with respect to adiabatic transport  times.
  Special attention is paid to shortcut-to-adiabaticity protocols that achieve, in
  faster-than-adiabatic times, the same results of slow adiabatic driving.
\end{abstract}

%
%
\maketitle
%
%
%
%
%
%
%
%
%
%
%
\section{Introduction}
Taking advantage of quantum phenomena to develop sensors, information processing, metrology,
or secure communications needs in general an exquisite control of the internal and/or motional state of
some quantum system. This control  is challenging,
because of environmental decoherence of superpositions,
and because the scale gap between  macroscopic control and microscopic
objects often leads to imperfect preparation or driving.
Thus, ``robustness'' of the manipulation  is a much needed quality
\cite{Deutsch2020}, even more so to scale up the control
to many qubits, as for quantum information processing \cite{Brucewicz2019}. Robustness is a multifaceted  concept,  relative to
initial conditions, noise, and loosely known or imperfectly controlled Hamiltonian parameters.
It is usually measured in terms of final quantum-state fidelities or acquired energy.

\vspace*{-.1cm}Ideas and techniques have been put forward to avoid or
mitigate the effects of noise, perturbations and imperfections, such as decoherence-free subspaces, error correction strategies,
or dissipation engineering \cite{Brucewicz2019}.
The ``process time'' is  also an important control element which affects robustness.
Times longer than the diabatic/adiabatic transition time scale  allow for adiabatic dynamics,  which are intrinsically robust with respect to smooth,
on-transit deviations of the control parameters. The trouble is that  slow processes also increase
decohering effects and excitations from noise. Moreover,
long adiabatic times may
not be compatible with
the
lifetime of the
system,
or with speeds desired for quantum information processing or for repetitions of sensing or algorithmic protocols.
Since shortcuts to adiabaticity (STA) \cite{Torrontegui2013,Guery2019} achieve the results of a slow adiabatic evolution in shorter times, they have attracted much attention. There are many shortcuts for a given process, allowing for
robustness optimization \cite{Ruschhaupt2012}.

\vspace*{-.03cm}  Different operations and systems
need adapted strategies towards robustness.
Here we focus on shuttling operations,  where a quantum system such as an atom, one or more ions,
cold-atom clouds, a Bose-Einstein condensate, or a magnon \cite{Kiely2021} is driven by a moving potential
a distance $d$ in a time
$t_f$.\footnote{For a particle of mass $m$ driven by harmonic trap of angular frequency $\omega$, the standard adiabaticity criterion
gives $t_f^2>>md^2 /(2\hbar\omega)$, namely, a characteristic kinetic energy smaller than the vibrational quantum \cite{Chen2011}.}
These are key manipulations for fundamental research and to implement
quantum-based technologies.\footnote{While our systems and motivation are ``quantum'', the shuttling itself may well be akin, or even identical  to ``classical'' transport, in particular for harmonic potentials.}
This Perspective provides an overview of theoretical and experimental efforts to
achieve STA-mediated robust transport, and of open questions and promising avenues.
STA has been reviewed recently with a broad scope \cite{Guery2019}, but  the peculiarities
of transport
should  benefit from a more specific
focus.  STA standard methods---invariant-based engineering, scaling, fast forward, or counterdiabatic driving---have emphasized simple analytical results, but treating   complex systems  typically entails numerical optimization. STA are understood here  as  a  continuum between the simple analytical results  and  numerical optimization of Optimal Control Theory techniques.

A seminal demonstration of STA-mediated diabatic transport used optical tweezers to transport cold atom clouds \cite{Couvert2008}
and deliver them in different experimental platforms. Individual atoms are also transported in optical-lattices
for on-demand positioning, quantum simulators, integration in photonic platforms, or to study  quantum random walks, see
\cite{Lu2020} and references therein.
Another example of a context in which robust transport is needed is ``STA-enhanced interferometry''
\cite{Navez2016,Dupont-Nivet2016,Palmero2017,Martinez-Garaot2018,Rodriguez-Prieto2020}, or more generally ``guided interferometry''
or sensors \cite{Ruster2017} based on moving  two branch potentials along some predetermined
separated trajectories
to measure e.g. rotations, unknown forces or fields, or to implement  quantum gates.
The interferometric phase is ``geometrical''  which provides robustness with respect to area preserving deviations
of the trajectories \cite{Rodriguez-Prieto2020}.

Further motivation is provided by the ``quantum charge-coupled device''  (QCCD) architecture
\cite{Wineland1998,Wan2020,Murali2020,Kaushal2020}, which aims at
making
quantum information processing with trapped ions scalable by keeping only small groups of ions in processing and storing
sites and shuttling ions among them diabatically \cite{Walther2012,Bowler2012}.

Even by narrowing down the scope of this review to robust shuttling, there is quite a broad landscape of systems,
settings
(e.g. with optical, magnetic, radio-frequency traps,  hybrid traps, or spin-chains), and potentials (lattices, quadratic, Gaussians  and others). Common
elements and concepts will be emphasized first, and specificities will be also pointed out later on. We assume an effective one-dimensional
description unless stated otherwise, and leave aside variants of the linear shuttling operations
such as turns at junctions, or separations and mergings.
\section{Different STA paradigms for fast particle shuttling}
There are several approaches to STA-mediated shuttling of a single particle or condensate in a trap, moving the  trap
center from $x_0(0)=0$ at initial time $t=0$ to $x_0(t_f)=d$ in a transport time $t_f$.
Assuming a rigid transport, without deformation, of a harmonic trap of (angular) frequency $\omega$, $x_0(t)$ may be inverse engineered
using quadratic invariants of motion \cite{Torrontegui2011} (alternatively ``scaling'' for condensates \cite{Muga2009,Schaff2011,Torrontegui2012}) or, equivalently, using methods based on the Fourier transform of the
trap velocity or acceleration \cite{Guery2014},
so that the excess energy of the final state compared to an adiabatic evolution vanishes regardless of the initial state of the particle.
This independence of the initial state is clearly a desirable robustness feature.
Virtually all treatments presuppose
a semiclassical driving Hamiltonian with $x_0(t)$ as a time-dependent ``external'' control not affected by the
transported particle because of the scale gap between the macroscopic control and the shuttled system,  a requisite for STA to be independent of the driven state \cite{Tobalina2018,Tobalina2019}.
The trajectory $x_0(t)$ is hardly unique. In the invariant approach, it is found  from a designed auxiliary classical trajectory $x_c(t)$ by solving inversely a forced-harmonic-oscillator equation,
$\ddot{x}_c(t)+\omega^2 x_c(t) =\omega^2 x_0(t)$. The only conditions to guarantee an excitation-free final state are
$x_c(0)=0$, $x_c(t_f)=d$ and that the time derivative $\dot{x_c}$
vanishes at boundary times. Other than that
there is full freedom to interpolate $x_c(t)$. (Higher derivatives may also be nullified to impose continuity
of the trap position and velocity at boundary times, which is usually desirable in practice).
This freedom,  characteristic of  STA methods, is a powerful asset to optimize robustness with respect to different
perturbations or errors. Invariant theory also provides explicit expressions for the dynamics when the frequency depends on time
\cite{Tobalina2017}, see also \cite{Hucul2008},
which are useful to calculate excitations and robustness with respect to such variations \cite{Lu2020}.
If both the frequency and the trap position are controllable, protocols may be designed inversely, e.g. to
launch a highly monochromatic wavepacket \cite{Tobalina2017}.

An alternative for excitation-free shuttling is to add control terms to some reference Hamiltonian
$H_0=p^2/(2m)+U[q-x_0(t)]$, which includes a rigidly transported potential $U$ not necessarily harmonic ($q$ and $p$ are position and momentum operators).
This addition can be done in different ways: The ``counterdiabatic (CD) approach''  applied to transport \cite{Torrontegui2011} assumes that the displacement of the initial  eigenvectors $|n(0)\ra$ of $H_0(0)$
according to $e^{-ipx_0(t)/\hbar} e^{-iE_nt/\hbar} |n(0)\ra$,  corresponds to an actual  dynamical evolution.
From the implied unitary evolution operator,
the driving Hamiltonian is deduced to be
$H_0+p\dot{x}_0$. Simple indeed, but it has not been realized so far, apart from a transport simulation in an interaction picture \cite{An2016}.
It
is in principle physically feasible in systems with actual or simulated spin-orbit coupling interactions $p\sigma_z$,
but note that the two spin components
suffer opposite forces, see \cite{Guery2019} and references therein.

Assuming a dynamical evolution  which shifts both the positions
and the momenta, i.e., a unitary evolution
$e^{im \dot{x}_0(t)q/\hbar} e^{-ipx_0(t)/\hbar}e^{-iE_nt/\hbar} |n(0)\ra$, implies instead the easier-to-realize driving Hamiltonian  $H=p^2/2m+U[q-x_0(t)]-mq\ddot{x}_0$. The added term compensates the inertial force. In a ``moving frame'', the interaction picture wavefunction  is that for a potential at rest. The compensating force  may as well be found  as a result of the ``fast forward approach'' \cite{Masuda2010}, ``invariant-based inverse engineering'' within a (Lewis-Leach \cite{Lewis1982}) family of Hamiltonians\footnote{A remarkable classical-quantum convergence occurs for the  Lewis-Leach Hamiltonians
%
$H={p^2}/({2m})-F(t)q+{m}\omega^{2}(t)q^{2}/2
+{\rho(t)^{-2}}U\!\left[({q-x_{0}(t)})/{\rho(t)}\right],
$
%
where
the scaling factor $\rho$ and the force $F$ satisfy Ermakov and Newton equations, which are common
for quantum or classical systems.
%
%
The quadratic invariant takes the same form, up to the correspondence $2qp\, {\rm (classical)}\sim qp+pq\, {\rm (quantum)}$, so classical and quantum systems can be inverse engineered in the same way
\cite{Gonzalez-Resines2017,Torrontegui2017,MartinezCercos2020,Lizuain2020}. STA approaches are as well applicable to stochastic processes described e.g. by Fokker-Planck equations \cite{Martinez2016}.} \cite{Torrontegui2011}, or by unitary transformation from the CD approach \cite{Ibanez2012}.
The main contribution of the compensating force to  ``robustness'' is that the trap can be arbitrary. As long as it is complemented by an adequate
linear term,
the details of the potential $U$, and specifically its anharmonicities, do not affect the excitation-free final result.
In interferometers using the compensating force,
the interferometric phase does not depend on the initial state, or on  perturbations that could affect the stability of the pivot point for the compensating potentials \cite{Rodriguez-Prieto2020}. Moreover,  the interferometric loop can be traversed quickly without adversely affecting sensitivity, and excitations that grow with process time can be mitigated.  (Other excitation types are possible as discussed below.)

Applying the compensating force is an elegant, simple solution for fast shuttling. However, the practical implementation of
the potential gradients depends on the system, experimental setting, transport distance and time.
To estimate feasibility, a lower bound from the maximum of  $\ddot{x}_0$ from the mean value theorem is $2d/t_f^2$.
For shuttling trapped ions in multisegmented
Paul traps control voltages are upper-bounded to avoid the breakdown of the insulating regions between the electrodes \cite{Furst2014,Tobalina2018}.  For shuttling neutral atoms in weaker optical or magnetic traps, achieving a proper compensation may be challenging  \cite{Torrontegui2011,Lam2021}. Approximations to the ideal compensation are also possible, for example, retaining only the harmonic part of $U[q-x_0(t)]$, the compensation amounts to implementing a shifted trajectory $x_0'(t)=x_0(t)+\ddot{x_0}/\omega^2$ \cite{Ibanez2012}. The approximation will fail for very fast processes as the particle escapes from the finite-depth trap. For  specific potential forms, such as a Gaussian for optical tweezers, this limitation
may be quantified in the form of  ``speed limit'' relations \cite{Sels2018}.

\vspace*{.22cm}The compensating method can be extended to condensates in the mean field regime \cite{Torrontegui2012} and formally to transport $N$ particles  \cite{Masuda2012}, but particles of different mass would need different forces \cite{Palmero2014}. For neutral atoms, different optical potentials could be set for the different species, whereas
within the current trapped-ion technology, forces are proportional to the charge, so the needed mass dependence of the forces is not feasible \cite{Palmero2014}.
And yet transporting different species together is of much interest. With a pair of different ions it is possible to sympathetically cool one species without perturbing the qubit in the other one, and furthermore, the use of ions of a different species as entanglement or measurement ancillas is widely considered to be necessary for large-scale quantum computing~\cite{Brucewicz2019}.  An approach to shuttle different-species ion pairs is to express the Hamiltonian in ``dynamical normal modes'' \cite{Palmero2014,Lizuain2017,Sagesser2020}, i.e., independent motions for time-dependent harmonic oscillators in the small-oscillations regime.
For a transport operation they can always be found by simple point transformations \cite{Lizuain2017}.  Invariant-based inverse engineering the trap trajectory can then be worked out  to satisfy the boundary conditions imposed on both independent oscillators for excitation-free transport. This needs in general  numerical optimization, but  analytical approximations perform satisfactorily \cite{Palmero2014} up to very short times for which the small-oscillation approximation breaks down.  Lu et al. \cite{Lu2015} used perturbation theory to design for pairs of different ions, trap trajectories
robust with respect to  slow frequency drifts in the scale of  the shuttling time.
For ions of equal mass in a harmonic trap the center of mass motion is separable, but not if the trap is anharmonic.
Palmero et al. \cite{Palmero2013} found  robust perturbation theory-based trap trajectories noticing that the dominant effect of a quartic anharmonicity is to shift the effective trap frequency for the center of mass.
\section{The ``Fourier method''}
If shuttling is performed in a harmonic oscillator---with constant frequency---and zero boundary conditions for the trap velocity,
the final excitation energy depends on the Fourier transform (FT) of the trap velocity or acceleration
at the trap frequency $\omega_0$
\cite{Reichle2006,Couvert2008,Bowler2012,Guery2014,MartinezCercos2020}. This was known for classical systems long ago \cite{Landau1976}, and it can be applied to the quantum case.
Fast transport which is excitation-free at the destination may thus
be achieved by trajectories that nullify the transform at $\omega_0$.

An important consequence
is that the final excitation effect of deviations from some ideal, excitation-free trajectory, only depends on the FT of the acceleration {\it deviation}. The excitation is formally independent of the ideal trajectory, and indeed in practice it will be independent when the deviation is itself independent, for example if it is due to background noise.
An indirect dependence may occur if the deviation is due to driving imperfections affected by the ideal trajectory.
Thus, smooth, band-limited ideal trajectories
should be preferable, although some settings allow for nearly-discontinuous, bang-bang protocols \cite{Alonso2013}.

Gu\'ery-Odelin  and Muga \cite{Guery2014}, see also \cite{MartinezCercos2020}, worked out a systematic method to
design trap trajectories that guarantee a vanishing transform at a chosen set of trap frequencies. This method may be useful for
multispecies transport, and also
to increase  the frequency window for which the transform vanishes around some $\omega_0$, thus allowing for robustness with
respect to errors in the actual trap frequency $\omega$ with respect to the nominal one $\omega_0$.
As $\omega$
is supposed constant, these trajectories are  of interest
when drift times of the trap frequency are larger than $t_f$.  In ion traps,
such variations can be caused by voltage drifts, thermal expansion, and charging of the trap surfaces \cite{Lu2014,Kaushal2020}.

Quite often robustness
with respect to one parameter comes with a price. In particular, increasing the frequency window width
implies
trap trajectories with stronger oscillations and also increasing spatial domains, and therefore higher transient energies.
These effects have been observed but not fully characterized \cite{Guery2014}.
Therefore, in a realistic scenario the window size cannot be increased
arbitrarily since the trap is effectively harmonic only up to a certain energy.
An open question is to what extent these broad-window trajectories are  robust with respect to
time-dependent
deviations of the frequency within the window, or anharmonicities. Since the trajectory design leaves room for different solutions, optimization strategies may be implemented.
\section{Noise in different parameters\label{nLu}}
Explicit analysis of the effect of noise in different parameters  on final energy excitation, using perturbative techniques
proposed in \cite{Ruschhaupt2012},
were worked out in Refs. \cite{Lu2014,Lu2015,Lu2018,Lu2020}. Here we focus on the work by Lu et al. \cite{Lu2020} as it covers the most complete
set of scenarios. Even if it falls short of addressing the broad span of complex noises that may be found in different settings,
it provides basic guidance and useful insights.
The trap is assumed to be an optical lattice
%
$
A \sin^2[Kx+\Phi],
$
%
in the deep lattice limit where anharmonicities may be neglected,
and for weak noise, which  allows the perturbative treatment. Random time-dependent fluctuations $\xi(t)$  may affect the three parameters  independently:  $K(t)=k[1+\lambda \xi(t)]$ (wavenumber ``accordion'' noise); $A(t)=a[1+\lambda\xi(t)]$  (amplitude or trap depth noise which amounts to a spring-constant noise);
and $\Phi=\phi(t)-\lambda\xi(t)$ (trap position noise), where $\lambda$ is the perturbative parameter. The fluctuating parameters are modeled by stochastic zero-mean and ``stationary'' deviations, i.e.,
the average over noise realizations  is ${\cal{E}}[\xi(t)]=0$, and the correlation function depends only on time difference, ${\cal{E}}[\xi(t)\xi(s)]=\alpha(t-s)$.

The ``noise sensitivities'' are the second order terms of the final energy expansion in powers of $\lambda$ and may be computed analytically
for specific noises, such as white noise, or Ohrnstein-Uhlenbeck (OU) colored noise, a simple model for examining non-zero correlation time effects.  It is the natural noise in a resistor-capacitor low-pass filter output voltage if noise in the
input voltage is white; it may also be used to produce flicker noise by combining different OU-noises \cite{Lu2014,Lu2018}.
The sensitivity can generally be split into ``static''---independent on the nominal (ideal) trap trajectory---and
``dynamical'' contributions that depend on the choice of ideal trajectory.
For fixed transport time only the later allow for some optimization.
An important finding is that position noise only leads
to static effects, so that the choice of nominal STA trap trajectory does not formally have any effect, consistent with the FT analysis in the previous section. In the white-noise limit the only way to decrease the
sensitivity for position noise is to shorten $t_f$. Colored OU noise implies sensitivity oscillations with respect to $t_f$
so that some time optimization is possible.
The sensitivities have different behavior with respect to $t_f$ for the different noisy parameters. In particular static and dynamical
sensitivities for spring-constant noise behave in opposite ways which leads to optimal transport times.
These optimal times  demonstrate that smaller process times are not necessarily the best strategy to
fight noise effects. It all depends on which parameter is noisy.

For transport times larger than the noise correlation time the excitation rates are proportional to the power spectral density at the trap frequency $\omega_0$ (noise in $\Phi$) or at
$2\omega_0$ (``parametric heating'' due to noise in $K$ or $A$). Changing the trap frequency may thus be worthwhile
if the shift diminishes the power spectral density.

Further research on the effects of noise and its mitigation may go along different lines, such
as making STA trajectories  ``noise resistant''  by imposing additional constraints on the invariant, e. g.  that its eigenbasis coincides
as much as possible with that of the noise operator \cite{Levy2018};
also, considering other noises (correlated for the different parameters, rather than independent, non-Markovian, position-dependent, ...);
optimizing trap trajectories to minimize dynamical effects;  or tracking the origin and form of the noises down to its  setting-dependent source. In a multisegmented Paul trap, for example,
the modeling could go more deeply by making the potential a function of electrode voltages and the circuits to produce them.
In the Paul trap, a quasi noise, e.g. in the trap frequency, may appear not due to actual electrode noise but to the numerical optimization to set the electrode voltages.
Further research to minimize and characterize this type of deviation and its effects is needed.
%
%
%
\section{Shortcuts to adiabaticity and Optimal control theory}
Shortcuts to adiabaticity based on invariants or inverse engineering approaches,
and Optimal Control Theory (OCT) blend quite well \cite{Chen2011,Torrontegui2012,Stefanatos2014_733,Furst2014,Mortensen2018,Amri2019,Ding2020,Zhang2021},
usually via Pontryagin's maximal principle.
While STA techniques provide families of ideal trap trajectories, OCT helps
to select the best among them to minimize some cost function, restricting the selection space if necessary to domains
imposed by physically motivated constraints.
Imposing constraints
is a way to make the transport ``robust'', as
problematic  domains are avoided. The constrained optimizations also set useful ``speed limits'', even if the
optimal trajectories found are hard to realize because of discontinuities.
For example, \cite{Chen2011} explores, indirectly, different ways to avoid the effect of anharmonicities by
finding, for harmonic traps,  minimal-time trajectories with constrained relative displacement
$|x_c-x_0|$, $x_c$ being the state center, as well as minimal (time-integrated) displacement trajectories for given time, and minimal (integrated)
potential energy trajectories for a given $t_f$. Also, Torrontegui et al \cite{Torrontegui2012} find minimal-time trajectories
that stay inside the domain $[0,d]$,  and Lau and Daniel minimize the Stark shift \cite{Lau2011} to set ultimate time limits of shuttling ion qubits because of dephasing and decoherence effects due to the driving field. In current ion-shuttling experiments these field effects are not significant \cite{Kaufmann2018}.
To avoid the practical difficulties implied by the sudden jumps in $x_0$ or its velocity, Ding et al. \cite{Ding2020}
work out smooth protocols adding further constraints on first and second derivatives of $x_c-x_0$, which imply larger shuttling times.


The anharmonicities may as well be included explicitly in the modeling and optimization:  Zhang et al. \cite{Zhang2016}, assuming  that the compensating force approach is not viable, minimize for fixed time
the time-averaged anharmonic potential energy for a potential with cubic  and quartic terms.
The strategy was to  work out the family of STA for a purely harmonic trap, and then combine perturbation theory and OCT  to choose the one that minimizes the perturbation. In \cite{Whitty2020} a different approach is proposed based on a gradient method, called ``enhanced shortcuts to adiabaticity'',
which provides trajectories which are not necessarily shortcuts for the harmonic problem.
Instead of these quantum treatments, in Refs. \cite{Zhang2015} \cite{Li2017} optimizations using  classical trajectories are performed.
Interestingly, ``magic times'' with very small excitation and remarkable robustness versus the parameter multiplying the anharmonic terms
are found \cite{Zhang2015}, which are approximately valid in corresponding quantum calculations.
Li et al. \cite{Li2017}, minimized the excitation
for some assumed trigonometric ansatzes and compared their performance.

Mortensen et al. \cite{Mortensen2018} suggest a generic strategy  to deal with robustness optimization that could be quite useful for shuttling.
They define a cost function that takes into account the sensitivity to perturbations
\cite{Ruschhaupt2012} as well as resource requirements. The optimization is performed among the STA control parameter
trajectories which guarantee
no excitation so it is less computationally expensive than other optimal control methods.
Simple cost functions can be quite effective to improve robustness \cite{Sagesser2020}, and different numerical optimization
subroutines or approaches may be used \cite{Stefanatos2020}.

OCT numerical  approaches maximizing the final fidelity or minimizing
the final energy without using invariant theory or inverse engineering as an intermediate step are also possible, see e.g. \cite{Furst2014,Amri2019,Lam2021}.
They share with  analytical STA methods the goal to achieve final excitation-free states via diabatic processes.
%
%
%
%
\section{Experiments and some system-specific aspects}
For a list of STA-mediated transport experiments see \cite{Guery2019}.
More recent experiments are described in Refs. \cite{Hickman2020,Pino2020,Kaushal2020,Lam2021,Lee2021}.
We comment here on works that have addressed robustness aspects in specific experimental settings.
Each setting involves particular three-dimensional potentials and a preliminary issue is to see if and how an effective
1D theory makes sense. This is often the case, but 3D effects may be relevant, in particular if the trap is expanded,  compressed \cite{Torrontegui2012a}, or rotated, and also due to nonlinearities.

The potential in a Gaussian-beam optical trap or lattice includes a radial-longitudinal coupling term \cite{Torrontegui2012a} which can be
made small but implies some effects.
In an approximate theory fitting well with experimental results for transport in a Gaussian-beam trap,
an averaging over radial direction shifts the trap frequency of the effective axial trap
\cite{Ness2018}. Ness et al. \cite{Ness2018}
also noted that trap trajectories  designed  via invariant-based inverse engineering
could have velocity discontinuities at the boundaries, but these discontinuities are  difficult to implement.
A solution is to use a higher order polynomial ansatz for  $x_c$.

Hickman et al. \cite{Hickman2020} work out a 1D theory for a moving lattice that takes into account that the transverse direction induces
a slow variation in trap frequency and trap depth. These effects are modeled by
applying  dephasing to the  density matrix with an empirical parameter.

The transport of Bose-Einstein condensates has also been studied with STA techniques for mean-field descriptions \cite{Schaff2011,Masuda2012,Torrontegui2012,Corgier2018,Amri2019}.
Three dimensional aspects are important because of the couplings implied  by the non-linear term, and non-trivial potential evolution  in some traps.
The experiments performed  in magnetic traps \cite{Schaff2011,Grzeschik2017} involve both transport and a simultaneous frequency change, and the
treatments and protocols are strongly trap dependent.
In  \cite{Schaff2011}, vertical transport is  a side effect of the combination of gravity and decompression of a Ioffe-Pritchard trap.
A generalization of invariant theory based on scaling laws in the Thomas-Fermi regime allowed to design a complete STA
protocol for radial and axial frequencies suppressing in principle  final excitations.
Measured  residual excitations were  attributed to several experimental
imperfections.  Ref. \cite{Grzeschik2017} deals with a dedicated transport experiment in an atom-chip trap with a $Z$-shaped configuration
in the absence of gravity. Transport is carried out in the $z$ direction, with  trap rotation in the $xy$ plane.
As the center-of-mass transport in the $z$ direction is uncoupled to the $xy$ directions, an approximate shortcut may be designed just for the
center of mass \cite{Corgier2018}, that takes into account cubic anharmonicities semiclassically,  and the $z$-dependence of the
frequency in that direction.
The results are shown to be robust with respect to timing errors and offsets in the magnetic bias field used to drive the trap
within current experimental resolutions. However,
since the shape of the condensate and the 3D couplings are disregarded some residual collective excitations
occur, even in ideal conditions. The condensate shape in the Thomas-Fermi regime could be taken into account
using OCT \cite{Amri2019},
and the extension of a full inverse engineered protocol as in \cite{Schaff2011} to this setting is an open question.

Henson et al. \cite{Henson2018}  implement a machine learning (ML)  algorithm to control the coupled decompression and transport of a
metastable helium condensate. After each experimental iteration
the excitation is measured and the algorithm adjusts its empirical model of the system to improve the control. This optimization
circumvents the need for an accurate theoretical modelling of a complex system. However,  black-box data-science models have their limitations too,
e.g. a dependence on spurious temporal relations, or the lack of a mechanistic understanding \cite{Karpatne2017}. A largely unexplored but promising
paradigm is a theory-guided data-science approach hybridizing the unique capabilities of machine learning with
theoretical knowledge of the underlying physics.

In linear Paul traps, the radial directions are decoupled from the axial potential  \cite{Blakestad2010}, but
transverse-axial couplings can occur due to non-ideal trap geometries, in particular for motionally excited states \cite{Kaushal2020}.
To shuttle ions in the axial direction, time-dependent voltages are applied to a set of DC trap electrodes.  The transport potential results from superposing the voltages and unit potentials of each trap electrode.
Once a desired transport potential is identified, solving for the electrode voltages to achieve such a potential is  generally an ill-posed inverse problem \cite{Blakestad2010, Blakestad2011, deClercq2015}. Several methods exist, including the pseudoinverse of a constraint matrix and least-squares minimization.  While the pseudoinverse method applied on a perfectly-matched problem (number of constraints equals the degrees of freedom) may result in voltage sets that exactly achieve the target potential, additional constraints such as voltage limits are unable to be included. This process for determining voltages can result in imperfect transport potentials.



To reduce high-frequency technical noise at the ions' locations, the trap electrodes are low-pass filtered, which results in output voltages that differ from inputs.  This imposes limitations if the electrode voltage ramps need frequencies above the cutoff \cite{Kaufmann2018}.  Accounting for technical constraints such as electrode filter response,  voltage limits, and slew rate requires optimization subroutines.  A machine learning approach to account for electrode filtering and systematic noise effects was used for optimizing voltages for rotating an ion crystal \cite{vanMourik2020}.  A similar method can be extended for shuttling. Other optimization techniques include feedback control to find optimal control parameters on segments of a transport operation and slow splitting and merging \cite{Bowler2015, Eble2010}. Local search techniques have also been used to optimize a robustness cost functional on the free parameters in a particular STA solution \cite{Mortensen2018,  Sagesser2020}.
Optimization on cost functions that include experimental data naturally lead to more robust shuttling operations, however full real-time feedback has not been implemented for trapped ions. While static experimental errors can be calibrated, time-dependent errors are not so simply accounted for. In this regard, further analysis of the relative effect of different time-dependent perturbations in transport controls would be useful.

Electric-field noise that heats ions from the motional ground state is a consideration for robust ion shuttling, as noise near the trap frequency will exist during protocols that maintain the curvature of the trap-potential well.  Noise originating on the trap-electrode surfaces~\cite{Brownnutt2015} is of current interest, as it can limit multi-qubit gate fidelity in small traps when technical noise sources are eliminated~\cite{Stuart2019}.  Some methods for electric-field-noise mitigation have been determined~\cite{KRBrown_arxiv_2020}, but a complete picture is lacking as the likely multiple underlying mechanisms are currently unknown~\cite{Sedlacek2018}. Protocols robust to mis-calibration of trap frequency can address noise on a timescale much longer than the transport time, but the ion is heated by noise close to the trap frequency, typically on the megahertz scale, faster than typically achievable transport times.  This will in general be separable from transport, with the trade off being that shuttling times as small as possible minimize heating from the surface while very short times lead to direct excitation of the ion motion.  For transport protocols in which the trap frequency is not maintained (``non-rigid'')~\cite{Sutherland_arxiv_2021}, parametric excitation must be taken into account, and additional surface-noise heating may accrue if the potential is weakened, due to the ion heating rate's roughly inverse-square dependence on the trap frequency~\cite{KRBrown_arxiv_2020}; 3D effects, as described above in the non-rigid case, are also a potential issue here.

Besides temporal variation of the trap frequency over slow and fast timescales as described above, spatial variation of the trap frequency along the transport path is also of concern in more complex ion-trap-array geometries.  The trap frequency as a function of distance along the trajectory can vary by more than $5$\% due to local trap imperfections, patches of charged insulator or adsorbate, or stray fields from other parts of the trap environment~\cite{Walther2012}; while these fields may be compensated at a sampling of points along the trajectory through calibration, this procedure adds overhead that scales with transport distance.  Robustness to potential-curvature variation is thus technologically beneficial for long trajectories in multi-electrode arrays.

In interferometers \cite{Martinez-Garaot2018,Rodriguez-Prieto2020}, or two-qubit gates \cite{Leibfried2003,Palmero2017},
ions may be subjected to hybrid traps, making use of the Paul trap to implement a stationary harmonic potential,
and  of a time- and  internal-state-dependent displacing forces caused by a gradient of detuned laser light.
Protocols with smooth field changes will favor robust operations. The intensity of the laser field might induce undesired internal state transitions,
and the scaling of this effect with process time was examined in \cite{Palmero2017}.

Fast transport of thousands of ions  has been also carried out and implies strong trap frequency
variations and anharmonicities \cite{Pedregosa-Gutierrez2015,Kamsap2015}. Transfer efficiencies are found to be quite sensitive to $t_f$ but in general the best transfer efficiencies are at short transport durations. There is room for improvement on the theoretical description to find
efficient and stable protocols for large ion clouds.
\section{Final comments}
We have pointed out existing and promising avenues towards
fast and safe shuttling of quantum particles. They are in general
``open loop'', without feedback control, easier to implement than closed loop approaches. However, open loop approaches suffer problems such as sensitivity to imperfect control functions, and rely on an accurate modeling of the systems. This suggests a concerted effort by theorists to achieve robust designs and by experimentalists to characterize systems and dominant control errors.
A further step would be to integrate these protocols into closed loop approaches based e.g.  on hybrid theory-guided machine learning.

\acknowledgments We thank I. Chuang, X. Chen, E. Torrontegui,  and D. Gu\'ery-Odelin for useful comments.
This work was supported by the Basque Country Government (Grant No. IT986-16)
and PGC2018-101355-B-I00 (MCIU/AEI/FEDER,UE). This material is based upon work supported by the Department of Defense and Under Secretary of Defense for Research and Engineering under Air Force Contract No. FA8702-15-D-0001. Any opinions, findings, conclusions or recommendations expressed in this material are those of the authors and do not necessarily reflect the views of the Department of Defense and Under Secretary of Defense for Research and Engineering.

%

\bibliography{Bibliography}{}

\providecommand{\href}[2]{#2}\begingroup\raggedright\begin{thebibliography}{10}

\bibitem{Deutsch2020}
I.~H. Deutsch, ``{\em Harnessing the Power of the Second Quantum Revolution}'',
  \href{http://dx.doi.org/10.1103/PRXQuantum.1.020101}{PRX Quantum {\bfseries
  1},  020101 (2020)}.

\bibitem{Brucewicz2019}
C.~D. Bruzewicz {\em et~al.}, ``{\em Trapped-ion quantum computing: Progress
  and challenges}'', \href{http://dx.doi.org/10.1063/1.5088164}{Appl. Phys.
  Rev. {\bfseries 6},  021314 (2019)}.

\bibitem{Torrontegui2013}
E.~Torrontegui {\em et~al.}, ``{\em {Shortcuts to Adiabaticity}}'',
  \href{http://dx.doi.org/10.1016/B978-0-12-408090-4.00002-5}{Adv. At. Mol.
  Opt. Phys. {\bfseries 62},  117--169 (2013)}.

\bibitem{Guery2019}
D.~Gu\'ery-Odelin {\em et~al.}, ``{\em Shortcuts to adiabaticity: Concepts,
  methods, and applications}'',
  \href{http://dx.doi.org/10.1103/RevModPhys.91.045001}{Rev. Mod. Phys.
  {\bfseries 91},  045001 (2019)}.

\bibitem{Ruschhaupt2012}
A.~Ruschhaupt {\em et~al.}, ``{\em {Optimally robust shortcuts to population
  inversion in two-level quantum systems}}'',
  \href{http://dx.doi.org/10.1088/1367-2630/14/9/093040}{New J. Phys.
  {\bfseries 14},  093040 (2012)}.

\bibitem{Kiely2021}
A.~Kiely {\em et~al.}, 2021.
\newblock pr

\bibitem{Chen2011}
X.~Chen {\em et~al.}, ``{\em {Optimal trajectories for efficient atomic
  transport without final excitation}}'',
  \href{http://dx.doi.org/10.1103/PhysRevA.84.043415}{Phys. Rev. A {\bfseries
  84},  043415 (2011)}.

\bibitem{Couvert2008}
A.~Couvert {\em et~al.}, ``{\em {Optimal transport of ultracold atoms in the
  non-adiabatic regime}}'',
  \href{http://dx.doi.org/10.1209/0295-5075/83/13001}{EPL {\bfseries 83},
  13001 (2008)}.

\bibitem{Lu2020}
X.-J. Lu {\em et~al.}, ``{\em Noise Sensitivities for an Atom Shuttled by a
  Moving Optical Lattice via Shortcuts to Adiabaticity}'',
  \href{http://dx.doi.org/10.3390/e22030262}{Entropy {\bfseries 22},  262
  (2020)}.

\bibitem{Navez2016}
P.~Navez {\em et~al.}, ``{\em Matter-wave interferometers using {TAAP}
  rings}'', \href{http://dx.doi.org/10.1088/1367-2630/18/7/075014}{New J. Phys.
  {\bfseries 18},  075014 (2016)}.

\bibitem{Dupont-Nivet2016}
M.~Dupont-Nivet {\em et~al.}, ``{\em Contrast and phase-shift of a trapped atom
  interferometer using a thermal ensemble with internal state labelling}'', New
  J. Phys. {\bfseries 18},  113012 (2016).

\bibitem{Palmero2017}
M.~Palmero {\em et~al.}, ``{\em {Fast phase gates with trapped ions}}'',
  \href{http://dx.doi.org/10.1103/PhysRevA.95.022328}{Phys. Rev. A {\bfseries
  95},  022328 (2017)}.

\bibitem{Martinez-Garaot2018}
S.~Mart\'{\i}nez-Garaot {\em et~al.}, ``{\em Interferometer with a driven
  trapped ion}'', \href{http://dx.doi.org/10.1103/PhysRevA.98.043622}{Phys.
  Rev. A {\bfseries 98},  043622 (2018)}.

\bibitem{Rodriguez-Prieto2020}
A.~Rodriguez-Prieto {\em et~al.}, ``{\em Interferometer for force measurement
  via a shortcut to adiabatic arm guiding}'',
  \href{http://dx.doi.org/10.1103/PhysRevResearch.2.023328}{Phys. Rev. Research
  {\bfseries 2},  023328 (2020)}.

\bibitem{Ruster2017}
T.~Ruster {\em et~al.}, ``{\em Entanglement-Based dc Magnetometry with
  Separated Ions}'', \href{http://dx.doi.org/10.1103/PhysRevX.7.031050}{Phys.
  Rev. X {\bfseries 7},  031050 (2017)}.

\bibitem{Wineland1998}
D.~J. Wineland {\em et~al.}

\bibitem{Wan2020}
Y.~Wan {\em et~al.}, ``{\em Quantum gate teleportation between separated qubits
  in a trapped-ion processor}'',.

\bibitem{Murali2020}
P.~Murali {\em et~al.}, ``{\em Architecting Noisy Intermediate-Scale Trapped
  Ion Quantum Computers}'', 2020 ACM/IEEE 47th Annual ISCA  529-542 (2020).

\bibitem{Kaushal2020}
V.~Kaushal {\em et~al.}, ``{\em Shuttling-based trapped-ion quantum information
  processing}'', \href{http://dx.doi.org/10.1116/1.5126186}{AVS Quantum Science
  {\bfseries 2},  014101 (2020)}.

\bibitem{Walther2012}
A.~Walther {\em et~al.}, ``{\em {Controlling Fast Transport of Cold Trapped
  Ions}}'', \href{http://dx.doi.org/10.1103/PhysRevLett.109.080501}{Phys. Rev.
  Lett. {\bfseries 109},  080501 (2012)}.

\bibitem{Bowler2012}
R.~Bowler {\em et~al.}, ``{\em {Coherent Diabatic Ion Transport and Separation
  in a Multizone Trap Array}}'',
  \href{http://dx.doi.org/10.1103/PhysRevLett.109.080502}{Phys. Rev. Lett.
  {\bfseries 109},  080502 (2012)}.

\bibitem{Torrontegui2011}
E.~Torrontegui {\em et~al.}, ``{\em {Fast atomic transport without vibrational
  heating}}'', \href{http://dx.doi.org/10.1103/PhysRevA.83.013415}{Phys. Rev. A
  {\bfseries 83},  013415 (2011)}.

\bibitem{Muga2009}
J.~G. Muga {\em et~al.}, ``{\em {Frictionless dynamics of Bose–Einstein
  condensates under fast trap variations}}'',
  \href{http://dx.doi.org/10.1088/0953-4075/42/24/241001}{J. Phys. B {\bfseries
  42},  241001 (2009)}.

\bibitem{Schaff2011}
J.-F. Schaff {\em et~al.}, ``{\em {Shortcut to adiabaticity for an interacting
  Bose-Einstein condensate}}'',
  \href{http://dx.doi.org/10.1209/0295-5075/93/23001}{EPL {\bfseries 93},
  23001 (2011)}.

\bibitem{Torrontegui2012}
E.~Torrontegui {\em et~al.}, ``{\em {Fast transport of Bose-Einstein
  condensates}}'', \href{http://dx.doi.org/10.1088/1367-2630/14/1/013031}{New
  J. Phys. {\bfseries 14},  013031 (2012)}.

\bibitem{Guery2014}
D.~Gu\'ery-Odelin {\em et~al.}, ``{\em Transport in a harmonic trap: Shortcuts
  to adiabaticity and robust protocols}'',
  \href{http://dx.doi.org/10.1103/PhysRevA.90.063425}{Phys. Rev. A {\bfseries
  90},  063425 (2014)}.

\bibitem{Tobalina2018}
A.~Tobalina {\em et~al.}, ``{\em Energy consumption for ion-transport in a
  segmented Paul trap}'', New J. Phys. {\bfseries 20},  065002 (2018).

\bibitem{Tobalina2019}
A.~Tobalina {\em et~al.}, ``{\em Vanishing efficiency of a speeded-up
  ion-in-Paul-trap Otto engine(a)}'',
  \href{http://dx.doi.org/10.1209/0295-5075/127/20005}{EPL {\bfseries 127},
  20005 (2019)}.

\bibitem{Tobalina2017}
A.~Tobalina {\em et~al.}, ``{\em {Fast atom transport and launching in a
  nonrigid trap}}'', \href{http://dx.doi.org/10.1038/s41598-017-05823-x}{Sci.
  Rep. {\bfseries 7},  5753 (2017)}.

\bibitem{Hucul2008}
D.~Hucul {\em et~al.}, ``{\em On the transport of atomic ions in linear and
  multidimensional ion trap arrays}'', Quantum Inf. Comput. {\bfseries 8},
  501--578 (2008).

\bibitem{An2016}
S.~An {\em et~al.}, ``{\em {Shortcuts to adiabaticity by counterdiabatic
  driving for trapped-ion displacement in phase space}}'',
  \href{http://dx.doi.org/10.1038/ncomms12999}{Nature {\bfseries 7},  12999
  (2016)}.

\bibitem{Masuda2010}
S.~Masuda {\em et~al.}, ``{\em {Fast-forward of adiabatic dynamics in quantum
  mechanics}}'', \href{http://dx.doi.org/10.1098/rspa.2009.0446}{Proc. R. Soc.
  Lond. A {\bfseries 466},  1135--1154 (2010)}.

\bibitem{Lewis1982}
H.~R. Lewis {\em et~al.}, ``{\em {A direct approach to finding exact invariants
  for one-dimensional time-dependent classical Hamiltonians}}'',
  \href{http://dx.doi.org/10.1063/1.525329}{J. Math. Phys. {\bfseries 23},
  2371 (1982)}.

\bibitem{Gonzalez-Resines2017}
S.~Gonz\'alez-Resines {\em et~al.}, ``{\em Invariant-Based Inverse Engineering
  of Crane Control Parameters}'',
  \href{http://dx.doi.org/10.1103/PhysRevApplied.8.054008}{Phys. Rev. Applied
  {\bfseries 8},  054008 (2017)}.

\bibitem{Torrontegui2017}
E.~Torrontegui {\em et~al.}, ``{\em Energy consumption for shortcuts to
  adiabaticity}'', \href{http://dx.doi.org/10.1103/PhysRevA.96.022133}{Phys.
  Rev. A {\bfseries 96},  022133 (2017)}.

\bibitem{MartinezCercos2020}
D.~Martínez-Cercós {\em et~al.}, ``{\em Robust load transport by an overhead
  crane with respect to cable length uncertainties}'',
  \href{http://dx.doi.org/10.1177/1077546319899204}{J. Vib. Control {\bfseries
  26},  1514-1522 (2020)}.

\bibitem{Lizuain2020}
I.~Lizuain {\em et~al.}, ``{\em Invariant-Based Inverse Engineering for Fast
  and Robust Load Transport in a Double Pendulum Bridge Crane}'',
  \href{http://dx.doi.org/10.3390/e22030350}{Entropy {\bfseries 22},  350
  (2020)}.

\bibitem{Martinez2016}
I.~A. Mart\'{\i}nez {\em et~al.}, ``{\em {Engineered swift equilibration of a
  Brownian particle}}'', \href{http://dx.doi.org/10.1038/nphys3758}{Nat. Phys.
  {\bfseries 12},  843-846 (2016)}.

\bibitem{Ibanez2012}
S.~Ib{\'{a}}{\~{n}}ez {\em et~al.}, ``{\em {Multiple Schr{\"{o}}dinger Pictures
  and Dynamics in Shortcuts to Adiabaticity}}'',
  \href{http://dx.doi.org/10.1103/PhysRevLett.109.100403}{Phys. Rev. Lett.
  {\bfseries 109},  100403 (2012)}.

\bibitem{Furst2014}
H.~A. F{\"{u}}rst {\em et~al.}, ``{\em {Controlling the transport of an ion:
  classical and quantum mechanical solutions}}'',
  \href{http://dx.doi.org/10.1088/1367-2630/16/7/075007}{New J. Phys.
  {\bfseries 16},  075007 (2014)}.

\bibitem{Lam2021}
M.~R. Lam {\em et~al.}, ``{\em Demonstration of quantum brachistochrones
  between distant states of an atom}'', Phys. Rev. X {\bfseries 11},  011035
  (2021).

\bibitem{Sels2018}
D.~Sels, ``{\em Stochastic gradient ascent outperforms gamers in the Quantum
  Moves game}'', \href{http://dx.doi.org/10.1103/PhysRevA.97.040302}{Phys. Rev.
  A {\bfseries 97},  040302 (2018)}.

\bibitem{Masuda2012}
S.~Masuda, ``{\em {Acceleration of adiabatic transport of interacting particles
  and rapid manipulations of a dilute Bose gas in the ground state}}'',
  \href{http://dx.doi.org/10.1103/PhysRevA.86.063624}{Phys. Rev. A {\bfseries
  86},  063624 (2012)}.

\bibitem{Palmero2014}
M.~Palmero {\em et~al.}, ``{\em {Fast transport of mixed-species ion chains
  within a Paul trap}}'',
  \href{http://dx.doi.org/10.1103/PhysRevA.90.053408}{Phys. Rev. A {\bfseries
  90},  053408 (2014)}.

\bibitem{Lizuain2017}
I.~Lizuain {\em et~al.}, ``{\em Dynamical normal modes for time-dependent
  Hamiltonians in two dimensions}'',
  \href{http://dx.doi.org/10.1103/PhysRevA.95.022130}{Phys. Rev. A {\bfseries
  95},  022130 (2017)}.

\bibitem{Sagesser2020}
T.~Sägesser {\em et~al.}, ``{\em Robust dynamical exchange cooling with
  trapped ions}'', \href{http://dx.doi.org/10.1088/1367-2630/ab9e32}{New J.
  Phys. {\bfseries 22},  073069 (2020)}.

\bibitem{Lu2015}
X.-J. Lu {\em et~al.}, ``{\em {Optimal transport of two ions under slow
  spring-constant drifts}}'',
  \href{http://dx.doi.org/10.1088/0031-8949/90/7/074038}{Physica Scripta
  {\bfseries 90},  074038 (2015)}.

\bibitem{Palmero2013}
M.~Palmero {\em et~al.}, ``{\em {Fast transport of two ions in an anharmonic
  trap}}'', \href{http://dx.doi.org/10.1103/PhysRevA.88.053423}{Phys. Rev. A
  {\bfseries 88},  053423 (2013)}.

\bibitem{Reichle2006}
R.~Reichle {\em et~al.}, ``{\em {Transport dynamics of single ions in segmented
  microstructured Paul trap arrays}}'',
  \href{http://dx.doi.org/10.1002/prop.200610326}{Fortschr. Phys. {\bfseries
  54},  666--685 (2006)}.

\bibitem{Landau1976}
L.~D. Landau and E.~M. Lifshitz, {\em Mechanics}.
\newblock Butterworth-Heinemann, 3~ed., 1976.

\bibitem{Alonso2013}
J.~Alonso {\em et~al.}, ``{\em Quantum control of the motional states of
  trapped ions through fast switching of trapping potentials}'', New J. Phys.
  {\bfseries 15},  023001 (2013).

\bibitem{Lu2014}
X.-J. Lu {\em et~al.}, ``{\em {Fast shuttling of a trapped ion in the presence
  of noise}}'', \href{http://dx.doi.org/10.1103/PhysRevA.89.063414}{Phys. Rev.
  A {\bfseries 89},  063414 (2014)}.

\bibitem{Lu2018}
X.-J. Lu {\em et~al.}, ``{\em Fast shuttling of a particle under weak
  spring-constant noise of the moving trap}'',
  \href{http://dx.doi.org/10.1103/PhysRevA.97.053402}{Phys. Rev. A {\bfseries
  97},  053402 (2018)}.

\bibitem{Levy2018}
A.~Levy {\em et~al.}, ``{\em Noise resistant quantum control using dynamical
  invariants}'', \href{http://dx.doi.org/10.1088/1367-2630/aaa9e5}{New Journal
  of Physics {\bfseries 20},  025006 (2018)}.

\bibitem{Stefanatos2014_733}
D.~{Stefanatos} {\em et~al.}, ``{\em Minimum-Time Quantum Transport With
  Bounded Trap Velocity}'',
  \href{http://dx.doi.org/10.1109/TAC.2013.2273296}{IEEE Trans. Autom. Control
  {\bfseries 59},  733-738 (2014)}.

\bibitem{Mortensen2018}
H.~L. Mortensen {\em et~al.}, ``{\em Fast state transfer in a $\Lambda$-system:
  a shortcut-to-adiabaticity approach to robust and resource optimized
  control}'', \href{http://dx.doi.org/10.1088/1367-2630/aaac8a}{New J. Phys.
  {\bfseries 20},  025009 (2018)}.

\bibitem{Amri2019}
S.~Amri {\em et~al.}, ``{\em Optimal control of the transport of Bose-Einstein
  condensates with atom chips}'',
  \href{http://dx.doi.org/10.1038/s41598-019-41784-z}{Sci. Rep. {\bfseries 9},
  (2019)}.

\bibitem{Ding2020}
Y.~Ding {\em et~al.}, ``{\em Smooth bang-bang shortcuts to adiabaticity for
  atomic transport in a moving harmonic trap}'',
  \href{http://dx.doi.org/10.1103/PhysRevA.101.063410}{Phys. Rev. A {\bfseries
  101},  063410 (2020)}.

\bibitem{Zhang2021}
Q.~Zhang {\em et~al.}, ``{\em Connection between Inverse Engineering and
  Optimal Control in Shortcuts to Adiabaticity}'',
  \href{http://dx.doi.org/10.3390/e23010084}{Entropy {\bfseries 23},  84
  (2021)}.

\bibitem{Lau2011}
H.-K. Lau and D.~F.~V. James, ``{\em Decoherence and dephasing errors caused by
  the dc Stark effect in rapid ion transport}'',
  \href{http://dx.doi.org/10.1103/PhysRevA.83.062330}{Phys. Rev. A {\bfseries
  83},  062330 (2011)}.

\bibitem{Kaufmann2018}
P.~Kaufmann {\em et~al.}, ``{\em High-Fidelity Preservation of Quantum
  Information During Trapped-Ion Transport}'',
  \href{http://dx.doi.org/10.1103/PhysRevLett.120.010501}{Phys. Rev. Lett.
  {\bfseries 120},  010501 (2018)}.

\bibitem{Zhang2016}
Q.~Zhang {\em et~al.}, ``{\em Optimal shortcuts for atomic transport in
  anharmonic traps}'',
  \href{http://dx.doi.org/10.1088/0953-4075/49/12/125503}{J. Phys. B {\bfseries
  49},  125503 (2016)}.

\bibitem{Whitty2020}
C.~Whitty {\em et~al.}, ``{\em Quantum control via enhanced shortcuts to
  adiabaticity}'',
  \href{http://dx.doi.org/10.1103/PhysRevResearch.2.023360}{Phys. Rev. Research
  {\bfseries 2},  023360 (2020)}.

\bibitem{Zhang2015}
Q.~Zhang {\em et~al.}, ``{\em Fast and optimal transport of atoms with
  nonharmonic traps}'',
  \href{http://dx.doi.org/10.1103/PhysRevA.92.043410}{Phys. Rev. A {\bfseries
  92},  043410 (2015)}.

\bibitem{Li2017}
J.~Li {\em et~al.}, ``{\em Trigonometric protocols for shortcuts to adiabatic
  transport of cold atoms in anharmonic traps}'',
  \href{http://dx.doi.org/https://doi.org/10.1016/j.physleta.2017.08.027}{Phys.
  Lett. A {\bfseries 381},  3272 - 3275 (2017)}.

\bibitem{Stefanatos2020}
D.~Stefanatos {\em et~al.}, ``{\em A shortcut tour of quantum control methods
  for modern quantum technologies}'',
  \href{http://dx.doi.org/10.1209/0295-5075/132/60001}{{EPL} {\bfseries 132},
  60001 (2020)}.

\bibitem{Hickman2020}
G.~T. Hickman {\em et~al.}, ``{\em Speed, retention loss, and motional heating
  of atoms in an optical conveyor belt}'',
  \href{http://dx.doi.org/10.1103/PhysRevA.101.063411}{Phys. Rev. A {\bfseries
  101},  063411 (2020)}.

\bibitem{Pino2020}
J.~M. Pino {\em et~al.}, 2020.
\newblock pr

\bibitem{Lee2021}
M.~Lee {\em et~al.}, ``{\em Ion shuttling method for long-range shuttling of
  trapped ions in {MEMS}-fabricated ion traps}'',
  \href{http://dx.doi.org/10.35848/1347-4065/abdabb}{Jpn. J. Appl. Phys.
  {\bfseries 60},  027004 (2021)}.

\bibitem{Torrontegui2012a}
E.~Torrontegui {\em et~al.}, ``{\em {Fast transitionless expansion of cold
  atoms in optical Gaussian-beam traps}}'',
  \href{http://dx.doi.org/10.1103/PhysRevA.85.033605}{Phys. Rev. A {\bfseries
  85},  033605 (2012)}.

\bibitem{Ness2018}
G.~Ness {\em et~al.}, ``{\em Realistic shortcuts to adiabaticity in optical
  transfer}'', \href{http://dx.doi.org/10.1088/1367-2630/aadcc1}{New J. Phys.
  {\bfseries 20},  095002 (2018)}.

\bibitem{Corgier2018}
R.~Corgier {\em et~al.}, ``{\em Fast manipulation of Bose{\textendash}Einstein
  condensates with an atom chip}'',
  \href{http://dx.doi.org/10.1088/1367-2630/aabdfc}{New J. Phys. {\bfseries
  20},  055002 (2018)}.

\bibitem{Grzeschik2017}
C.~Grzeschik, 2017.
\newblock {P}h.D thesis Humboldt-Universit{\"a}t zu Berlin.

\bibitem{Henson2018}
B.~M. Henson {\em et~al.}, ``{\em Approaching the adiabatic timescale with
  machine learning}'',.

\bibitem{Karpatne2017}
A.~Karpatne {\em et~al.}, ``{\em Theory-Guided Data Science: A New Paradigm for
  Scientific Discovery from Data}'', IEEE Trans. Knowl. Data Eng. {\bfseries
  29},  2318-2331 (2017).

\bibitem{Blakestad2010}
B.~Blakestad, 2010.
\newblock {P}h.D. thesis University of Colorado.

\bibitem{Blakestad2011}
R.~B. Blakestad {\em et~al.}, ``{\em Near-ground-state transport of trapped-ion
  qubits through a multidimensional array}'',
  \href{http://dx.doi.org/10.1103/PhysRevA.84.032314}{Phys. Rev. A {\bfseries
  84},  032314 (2011)}.

\bibitem{deClercq2015}
L.~E. de~Clercq, 2015.
\newblock {P}h.D. thesis ETH Z\"{u}rich.

\bibitem{vanMourik2020}
M.~W. van Mourik {\em et~al.}, ``{\em Coherent rotations of qubits within a
  surface ion-trap quantum computer}'',
  \href{http://dx.doi.org/10.1103/PhysRevA.102.022611}{Phys. Rev. A {\bfseries
  102},  022611 (2020)}.

\bibitem{Bowler2015}
R.~Bowler, 2015.
\newblock {P}h.D. thesis University of Colorado.

\bibitem{Eble2010}
J.~F. Eble {\em et~al.}, ``{\em Feedback-optimized operations with linear ion
  crystals}'', \href{http://dx.doi.org/10.1364/josab.27.000a99}{J. Opt. Soc.
  Am. B {\bfseries 27},  A99 (2010)}.

\bibitem{Brownnutt2015}
M.~Brownnutt {\em et~al.}, ``{\em Ion-trap measurements of electric-field noise
  near surfaces}'', \href{http://dx.doi.org/10.1103/revmodphys.87.1419}{Rev.
  Mod. Phys. {\bfseries 87},  1419–1482 (2015)}.

\bibitem{Stuart2019}
J.~Stuart {\em et~al.}, ``{\em Chip-Integrated Voltage Sources for Control of
  Trapped Ions}'',
  \href{http://dx.doi.org/10.1103/PhysRevApplied.11.024010}{Phys. Rev. Applied
  {\bfseries 11},  024010 (2019)}.

\bibitem{KRBrown_arxiv_2020}
K.~R. Brown {\em et~al.}, 2020.
\newblock pr

\bibitem{Sedlacek2018}
J.~A. Sedlacek {\em et~al.}, ``{\em Evidence for multiple mechanisms underlying
  surface electric-field noise in ion traps}'',
  \href{http://dx.doi.org/10.1103/PhysRevA.98.063430}{Phys. Rev. A {\bfseries
  98},  063430 (2018)}.

\bibitem{Sutherland_arxiv_2021}
R.~T. Sutherland {\em et~al.}, 2021.
\newblock pr

\bibitem{Leibfried2003}
D.~Leibfried {\em et~al.}, ``{\em {Experimental demonstration of a robust,
  high-fidelity geometric two ion-qubit phase gate.}}'',
  \href{http://dx.doi.org/10.1038/nature01492}{Nature {\bfseries 422},  412--5
  (2003)}.

\bibitem{Pedregosa-Gutierrez2015}
J.~Pedregosa-Gutierrez {\em et~al.}, ``{\em {Ion transport in macroscopic RF
  linear traps}}'', \href{http://dx.doi.org/10.1016/j.ijms.2015.03.008}{Int. J.
  Mass Spectr. {\bfseries 381-382},  33--40 (2015)}.

\bibitem{Kamsap2015}
M.~R. Kamsap {\em et~al.}, ``{\em Fast and efficient transport of large ion
  clouds}'', \href{http://dx.doi.org/10.1103/PhysRevA.92.043416}{Phys. Rev. A
  {\bfseries 92},  043416 (2015)}.

\end{thebibliography}\endgroup
\bibliographystyle{sofia}
%
%
%
%

%
%
%
%
%

\end{document}